\documentclass[12pt]{iopart}

\usepackage{iopams}
 
\def\IC{\relax\,\hbox{$\inbar\kern-.3em{\rm C}$}}
\def\bfzero{\relax\,\hbox{$\inbar\kern-.3em{\rm 0}$}}

\def\bfone{\relax{\rm 1\kern-.35em 1}}

 \def\cG{{\cal G}}

\def\cN{{\cal N}}

\def\beq{\begin{equation}}
\def\eeq{\end{equation}}
\def\bea{\begin{eqnarray}}
\def\eea{\end{eqnarray}}
\def\bet{\begin{tabular}}
\def\eet{\end{tabular}}
\def\bes{\begin{subequations}\bea}
\def\ees{\eea\end{subequations}}
\newcommand{\dfrac}{\displaystyle \frac}

\def\a{\alpha}

\def\da{\dot{\alpha}}

\def\e{\epsilon}

\def\IC{\hbox{\msbm C}}

\begin{document}
\jl{6}

\begin{flushright}
 CERN--TH/99-304 \\
 DFTT 51/99
\end{flushright}

\title[Superconformal field theories from IIB spectroscopy]
{Superconformal field theories from IIB spectroscopy on $AdS_5\times T^{11}$}

\author{A  Ceresole\dag\footnote[2]{Permanent Address:  Dipartimento di
Fisica, Politecnico di Torino, Corso Duca Degli Abruzzi 24, I-10129 Torino,
Italy.}
, G Dall'Agata\S , R D'Auria$\|$ and
S Ferrara\dag}

\address{\dag\  TH Division, CERN, 1211 Geneva 23, Switzerland }
\address{\S\  Dipartimento di Fisica Teorica, Universit\`a  di Torino
 via P. Giuria 1, I-10125 Torino, Italy}

\address{$\|$\ Dipartimento di Fisica, Politecnico di Torino,
C.so Duca degli Abruzzi, 24, I-10129 Torino, Italy}

\begin{abstract}

We report on  tests of the AdS/CFT correspondence
that are made possible by complete knowledge of the
Kaluza--Klein mass spectrum of type IIB supergravity
on $AdS_5 \times T^{11}$ with $T^{11}=SU(2)^2/U(1)$.
After briefly discussing general multiplet shortening
conditions in $SU(2,2|1)$ and $PSU(2,2|4)$ ,
we compare various types of short  $SU(2,2|1)$
supermultiplets on $AdS_5$
and different families of boundary operators with protected
dimensions. The supergravity analysis  predicts the occurrence in the
SCFT at leading order in $N$ and $g_s N$, of extra towers of long multiplets
whose dimensions are rational but not protected by supersymmetry.

\end{abstract}


\bigskip

\vspace{3cm}

\bigskip

{\it To appear in the proceedings of the } STRINGS '99  {\it conference,
Potsdam (Germany), 19--25 July 1999}

\noindent
CERN--TH/99-304

\maketitle

\section{Introduction}

One of the most stringent checks on  the AdS/CFT correspondence \cite{M,GKP,W}
is the matching between the mass spectrum of the Kaluza--Klein (KK)
supergravity models and the conformal dimensions of superconformal primary
operators of the boundary superconformal field theory.
This probes the correspondence at least  in the regime where $g_s N$
($g_s$ being the string coupling) and/or $N$ are  large. After the strong
support provided by tests made for maximal
supersymmetry, where the dynamics of $N$ coincident D3 branes
(for large $N$) is related to type IIB supergravity compactified
on $AdS_5\times S_5$ \cite{AdSReview}, it is natural to consider
lower supersymmetry, where a far richer structure
of matter  multiplets leads to additional symmetries beside the original
$R$-symmetry.
Alternatively to orbifolding the sphere $S_5$,
a very interesting  way to reduce supersymmetry is to consider coset models
such as $ T^{pq}=\dfrac{SU(2)\times SU(2)}{U(1)}$
($p$ and $q$ define the embedding of the $H=U(1)$ group into the two
$SU(2)$ groups), which yield for $p=q=1$ an $\cN=2$ supergravity theory with a matter
gauge group $G=SU(2)\times SU(2)$ \cite{Rom}.
The corresponding $CFT_4$ description was constructed in
\cite{KW,MP} as an $\cN=1$
Yang--Mills theory with a flavour symmetry $G$.
One finds a conformal
field theory with ``singleton'' degrees of freedom $A$
and $B$, each a doublet of the  factor groups $SU(2)\times SU(2)$
and with conformal anomalous dimension $\Delta_{A,B}=3/4$.  The
gauge group $\cG$ is $SU(N)\times SU(N)$ and the two singleton
(chiral) multiplets are respectively in the $(N,\overline N)$ and
$(\overline N,N)$ of $\cG$.
The gauge potentials lying in the adjoint of one of the two $SU(N)$
groups, whose field--strength in superfield notation is given by $W_\alpha$,
are singlet of the matter groups, carry unit $U_R(1)$ charge and have $\Delta = 3/2$.

There is also a superpotential  \cite{KW} $V = \lambda \e^{ij} \e^{kl} Tr(A_i B_k A_j
B_l)$ with $\Delta = 3$, $r = 2$ playing an important role in the
discussion, since it determines to some extent both the chiral
spectrum and the marginal deformations of the SCFT.

Chiral operators which are the analogue of
the KK excitations of the $SU(N)$ $\cN=4$ Yang--Mills theory
is given by ${\it Tr}(AB)^k$ with $R$--charge $k$ and in the
($\frac{k}{2}$,$\frac{k}{2}$) representation of $SU(2)\times SU(2)$ \cite{KW}.

More generally \cite{CDDF}, there exist a complete correspondence between
all the CFT operators and the KK modes for the conformal operators of
preserved scaling dimension. Even
more intriguingly,
there exist other operators related to long
multiplets but having nonetheless non--renormalised conformal dimension
in the large $N$, $g_s N$ limit.
These seem to be the lowest dimensional
ones for a given structure appearing in the supersymmetric Born--Infeld
action of the $D3$--brane on $AdS_5\times T^{11}$ \cite{ts}.

It is now well known that states that are  associated to
{\it shortened} multiplets of the superconformal algebra
$SU(2,2|{\cal N})$ for the $AdS_5$/$CFT_4$ duality, in virtue of supersymmetry
have protected conformal dimensions. They  are
BPS states from the point of view of the bulk theory and shortened superfields
from the boundary perspective.

After this introduction, we present a brief review of the
multiplet shortening conditions for the $SU(2,2|1)$ and
$PSU(2,2|4)$ superalgebras, while in section 3 we show the results
of the comparison between the dual $AdS$ and CFT theories for the
$T^{11}$ example.

\section{Group theory lore: UIR's of $SU(2,2|1)$ and $PSU(2,2|4)$ }

We first consider the unitarity bounds for the
highest-weight representations of the $SU(2,2|\cN)$ superalgebra in
the $\cN=1$ and $\cN=4$ cases, that are those relevant for the analysis
of the $T^{11}$ and the $S_5$  IIB compactifications respectively .

For the $SU(2,2)$ algebra itself, a given UIR is denoted,
following Flato and Fr\o nsdal \cite{FF}, as
$D(E_0,J_1,J_2)$ where $E_0,J_1,J_2$ are the quantum numbers of the
highest-weight state, given by a finite UIR of the maximal
compact subgroup $SU(2)\times SU(2)\times U(1)$. The UIR's fall in
three series \cite{BFH},
\bea
\label{bound}
a)\, J_1J_2\ne 0\qquad\qquad &E_0\ge 2+J_1+J_2\\
b)\, J_2J_1=0\qquad\qquad &E_0\ge 1+J\\
c)\, J_1=J_2=0\qquad\qquad& E_0=0\ .
\eea
In the bulk interpretation, the inequalities in a) and b) yield
massive $AdS_5$ representations.  Their saturation
gives rise  to massless  particles of {\it spin} $J_1+J_2$
in the case a) and to singletons of spin $J$ for the b) threshold.

Note that in the AdS/CFT map the bulk-boundary quantum
numbers $(E_0,J_1,J_2)$ refer to the {\it compact} basis for the
AdS states, while they refer to the non-compact basis
$SL(2,C)\times O(1,1)$ for the boundary conformal operators
\cite{W,GMZtwo}. The highest weight state in AdS is related
to a conformal operator $O(x)$ at $x=0$, and thus the AdS energy $E_0$
becomes  the conformal dimension $\Delta_0$ while the
$(J_1,J_2)$ labels give the Lorentz spin of
$O(x)$.

From the CFT perspective,  the threshold value for the bound a)
represents a conformal conserved currents of spin $J=J_1+J_2$,
\beq
\label{boundII}
E_0=2+J_1+J_2\quad(J_1J_2\ne
0)\quad\rightarrow\quad
\partial^{\alpha_1\dot\alpha_1}J_{\alpha_1...\alpha_{2J_1},
\dot\alpha_1...\dot\alpha_{2J_2}}(x)=0,
\eeq
while for the bound b) one gets  massless spin $J$ conformal
fields on the boundary,
\bea
\label{boundIItwo}
E_0=1+J\quad\qquad (J\ne 0)\qquad\qquad \rightarrow \partial^{\alpha_1\dot\alpha_1}
O_{\alpha_1...\alpha_{2J}}=0\\
\phantom{E_0=1+J}\quad\qquad (J=0)\qquad\qquad\rightarrow\partial^2 O(x)=0
\nonumber      .
\eea
The case c) gives rise  to the identity representation.

Generalising to the $SU(2,2|\cN)$ superalgebras \cite{BG,DP}, the
highest weight state is denoted by
$D(E_0,J_1,J_2;r,a_1,...,a_{\cN-1})$, where the quantum numbers in
brackets indicates an UIR of $SU(2,2)\times U(1)\times SU(\cN)$,
$r$ lebelling the $U(1)$ R-symmetry and
$a_1,...,a_{\cN-1}$ the Dynkin labels of a UIR of the non-abelian
symmetry $SU(\cN)$. We will denote by R the $U(1)$ generator inside $U(\cN)$.

Note that for $\cN\ne 4$, the $SU(2,2|\cN)$ algebra is both a
subalgebra and a quotient algebra of $U(2,2|\cN)$, since the
supertrace generator (which is a central charge) can be eliminated
by a redefinition of the  R generator.
However, this redefinition is not  possible for
$\cN=4$ since inthat case $R$ drops from the supersymmetry anti-commutators and
becomes an outer automorphism of the algebra \cite{B}.

Therefore there are two inequivalent algebras (which do
not include the  R generator), $PSU(2,2|4)$ and $PU(2,2|4)$,
depending on whether $r=0$ or $r\ne 0$ (for $\cN = 4$, $r$ denotes
the central charge.

Since the $\cN=4$ Yang--Mills multiplet has $r=0$, we will only
consider $PSU(2,2|4)$. In the boundary CFT language, where UIR's can be
realized as conformal superfields, the superhighest weight state
corresponds to a superfield $\phi(x,\theta)$ at $x=\theta=0$
\cite{DP,AF,FZ,PARK}.

The unitarity bounds for $SU(2,2|1)$ were given in \cite{FF,DP,FGPW}.
They generalize the cases a), b) and c) of eq.
(\ref{bound}) and read
\beq
\!\!\!\!\!\!\!\!\!\! A) \, E_0\ge 2+2J_2+\frac{3}{2}r\ge 2+2J_1-\frac{3}{2}r\qquad
(\hbox{or}\, J_1\rightarrow J_2,\, r\rightarrow
-r)\quad J_1,J_2\ge 0
\eeq
which implies
\beq
E_0\ge
2+J_1+J_2,\qquad \frac{3}{2}r\ge J_1-J_2,\qquad 2+2J_1-E_0\le \frac{3}{2}r\le
E_0-2-2J_2.
\eeq

\beq
\!\!\!\!\!\!\!\!\!\! B)\, E_0=\frac{3}{2}r\ge
2+2J-\frac{3}{2}r\qquad (J_2=0,\,J_1=J,\, \hbox{or} \, J_1=0,\,J_2=J,\,
r\rightarrow -r)
\eeq
and thus $E_0\ge 1+J$. Finally
\beq
C) \, E_0=J_1=J_2=r=0
\eeq
which is the identity representation.

Shortening in the case A) takes place when
\beq
\label{shorta}
E_0=2+2J_2+\frac{3}{2}r,\qquad \left(\frac{3}{2}r\ge J_1-J_2\right)
(\hbox{or}\, J_1\rightarrow J_2,\, r \rightarrow -r)\ .
\eeq
This is a semi-long AdS$_5$ multiplet or, in conformal language, a
{\it semiconserved} superfield \cite{CDDF,O},
\beq
\label{op}
\bar
D^{\dot\alpha_1}
L_{\alpha_1...\alpha_{2J_1},
\dot\alpha_1...\dot\alpha_{2J_2}}
(x,\theta,\bar\theta)=0,\qquad
(\bar D^2L_{\alpha_1...\alpha_{2J_1}}=0\, \hbox{for}\, J_2=0)
\eeq
(in our conventions $\theta$ carries $\Delta=-1/2,r=1,\bar\theta$ has
$\Delta=-1/2,r=-1$).

Maximal shortening for the bound A) happens for $E_0=2+J_1+J_2,\,
r=J_1-J_2$. This is a conserved superfield which satisfies both
left and right constraints:
\beq
\label{cons}
 \bar D^{\dot\alpha_1}J_{\alpha_1...\alpha_{2J_1},
\dot\alpha_1...\dot\alpha_{2J_2}}=
D^{\alpha_1}J_{\alpha_1...\alpha_{2J_1},\dot\alpha_1...\dot\alpha_{2J_2}}=0
\eeq

Further,  shortening in B) corresponds to {\it chiral superfields} $r = 2/3 E_0$,
while maximal shortening  to {\it massless} chiral superfields,
i.e. chiral singleton representations: $E_0=\frac{3}{2}r=1+J$. The
superfield, for $E_0=\frac{3}{2}r$ satisfies,
\beq
\label{chiral}
\bar D^{\dot\alpha}S_{\alpha_1...\alpha_{2J}}=0
\eeq
and, for $E_0=1+J$, it also satisfies
\beq
\label{chiraltwo}
D^{\alpha_1}S_{\alpha_1...\alpha_{2J}}=0\qquad (D^2 S=0,\, \hbox{for}\, J=0)
\eeq
These equations are the supersymmetric
version of (\ref{boundII}) and (\ref{boundIItwo}).

With an abuse of language, we may call off-shell singletons
chiral superfields since in an interacting conformal
field theory singletons may acquire anomalous dimension,  and thus fall
in (\ref{chiral}).

It is also evident, from superfield multiplication, that by taking
suitable products of several free supersingletons one may
get any other superfield of type (\ref{op}), (\ref{cons}) or (\ref{chiral}).

One can remark that, since the shortening condition just implies a relation
between $E_0$ and $r$ without fixing their value,
superfields obeying (\ref{op}), (\ref{chiral}) may have anomalous
dimensions .

The basic singleton multiplets for $\cN=1$ gauge theories arise
for $J=0,1/2$ in (\ref{chiral}), i.e. chiral scalar superfields $S$
(Wess-Zumino multiplets) and Yang-Mills field strength multiplets
$W_\alpha$. Any other conformal operator is obtained by suitable
multiplication of these two sets of basic superfields.

In type IIB supergravity on $T^{11}$ long, semi-long and chiral multiplets do
indeed occur \cite{KW,G,CDDF}.
Chiral WZ singleton multiplets
have in this case an anomalous dimension $\gamma=-1/4$
($\Delta=1+\gamma$) and R-symmetry $r=3/4$.

The $\cN=4$ superalgebra is of great interest because it corresponds
to $\cN=4$ superconformal Yang-Mills theory and lives, in the dual
description, at the boundary of AdS$_5$
\cite{M,GKP,W}.
The supergravity theory emerges as the low
energy limit of type IIB string theory compactified on
AdS$_5\times$S$_5$.

The highest weight UIR's of the $PSU(2,2|4)$ superalgebra are
denoted by $D(E_0,J_1,J_2;p,k,q)$, where $(p,k,q)$ are the $SU(4)$
Dynkin labels.

There exist three classes of UIR's
\beq
\label{superboundAprime}
A^\prime)\qquad E_0\ge
2+J_1+J_2+p+k+q,\qquad J_2-J_1\ge {1\over
2}(p-q),
\eeq
with
 maximal shortening occurring when,
\beq
\label{maxshortenAprime}
E_0=2+J_1+J_2+p+k+q,\qquad J_2-J_1={1\over2}(p-q).
\eeq
Massless bulk multiplets arise for $p=k=q=0$ and
$J_1=J_2$.
\beq
\label{superboundBprime}
B^\prime)\qquad E_0={1\over 2}(p+2k+3q)\ge 2+2J+{1\over 2}(3p+2k+q)
\eeq
$$
(\hbox{$J_2=0$,$J_1=J$ or $J_1\rightarrow J_2$,
$(p,k,q)\rightarrow (q,k,p)$})
$$
 with
\beq
\label{conseqBprime}
E_0\ge 1+J+p+k+q\qquad 1+J\le {1\over 2}(q-p).
\eeq
Maximal shortening occurs when $1+J= {1\over 2}(q-p)$, with
highest weight $D(3+3J+2p+k,J,0;p,k,p+2+2J)$. No supersingletons
appear in this series. Finally,
\beq
\label{superboundCprime}
C^\prime)\qquad
E_0=2p+k,\qquad p=q,\qquad J_1=J_2=0.
\eeq
The highest weight states are $D(2p+k,0,0;p,k,p)$. The $p=0, k\ge 2$
UIR's correspond to the KK states of type IIB on AdS$_5\times
S_5$, the $k=2$ case being associated with the bulk graviton
multiplet. The $p=0,k=1$ UIR yields the only
supersingleton of the $PSU(2,2|4)$ algebra \cite{B,GMZone}. The
infinite sequence of UIR's with $p=0$, multiplets with
$J_{MAX}=2$ have been obtained in \cite{GM} with the oscillator
construction. They are associated with  the harmonic \label{GIKO} holomorphic
superfields of \cite{HW}.
The case $p\ne 0$, which may be relevant for multiparticles supergravity
states, has been discussed in \cite{zafser}.

\section{Confronting with experiment}

$T^{pq}=SU(2)^2/U(1)$ cosets are Einstein spaces having
$\cN=2$ supersymmetry only when the subgroup $U(1)$
generator $T_H=p \sigma_3 +q \hat \sigma_3$ is defined with $p=q=1$,
where $\sigma_3, \hat \sigma_3$ are Pauli matrices generating the two $SU(2)$
groups in the numerator.
The $U(1)$ $R$-symmetry generator is $T_R=\sigma_3-\hat\sigma_3$.
Since $T^{11}$ is topologically the product
$S_2 \times S_3$, it has non--trivial Betti numbers
$b_2=b_3=1$, and therefore
the full isometry group is $SU(2,2|1)\times SU(2)\times SU(2)$
with an extra $U_B(1)$ gauge symmetry related to the existence of
non--trivial three cycles \cite{GKL}.

Knowing the fundamental degrees of freedom of the conformal
field theory, one could try to write the conformal operators
by simply combining the above fields while respecting the
symmetries of the theory.
Next to the already mentioned  $Tr(AB)^k$ chiral primaries, one
could also have an operator given by $Tr[W_\alpha (AB)^k]$ or $Tr[W^2
(AB)^k]$, and so on.
The important point is that the correspondence with the
KK states is true only for the protected operators, and
thus one needs to know  these latters to make the comparison.

\subsection{CFT $\to$ AdS}

The operators with protected conformal dimension correspond
to the short representation of the $SU(2,2|1)$ supergroup
described in the  previous section.
In our case we have only three types of such operators,
namely the {\it chiral} (\ref{chiral}), {\it conserved} (\ref{cons})
and {\it semi--conserved} (\ref{op})  superfields.
Since these fields satisfy certain specific
constraints effecting their quantum numbers, their
anomalous dimension is also fixed in terms of their spin and
$R$--symmetry charge.

It is easy to relate operators of different type by superfield
multiplication. The product of a chiral $(J_1,0)$ and an anti--chiral $(0,J_2)$
primary gives a generic superfield with
$(J_1,J_2)$, $\Delta = \Delta^c +
\Delta^a$ and $r = \frac{2}{3}(\Delta^c - \Delta^a)$.
By multiplying a {\it conserved current} superfield $J_{\a_1 \ldots
\a_{2J_1}, \da_1 \ldots \da_{2J_2}}$ by a chiral scalar
superfield one gets a semi--conserved superfield with $\Delta =
\Delta^{c} + 2 + J_1 + J_2$  and $r = \frac{2}{3}(\Delta-2-2J_2)$.

These are the basic rules to construct operators with protected
dimensions beside the chiral ones, and they also apply in superconformal
field theories of lower or higher dimensions. For instance, beyond the
chiral operators with $\Delta = r$ (hypermultiplets), in $d = 3$ $OSp(2|4)$
superconformal field theories one replaces the shortening condition
(\ref{op}) by the simpler constraint
\bea
D^{-\alpha_1} L_{\a_1 \ldots \a_{2s}} (x,\theta,\bar{\theta}) &=& 0
\qquad\qquad s \neq 0 \\
D^{-}{}^2 L (x,\theta,\bar{\theta}) &=& 0
\qquad\qquad s = 0 ,
\eea
defining semiconserved tensor operators ($s = 0,\frac{1}{2},1$ in the KK context) with
protected dimensions $\Delta = 1 + s + r$ (or $r \to -r$ if $-
\to +$).
As before, $L$ is obtained by multiplying a conserved spin $s$
superfield (for which both $D^+$ and $D^-$ constraints are
satisfied) and a chiral superfield\footnote{
The above basic rules have been recently used in the AdS/CFT
correspondence of  some $M$--theory models compactified on AdS$_4 \times
X_7$ \cite{Zaff}.}.

Since the anomalous dimensions of these operators is fixed in terms of
their spin and $R$--symmetry, it must be given by rational number. This
yields a very restrictive condition when searching for the corresponding
supergravity states, as it  imposes strong constraints on
the allowed masses and matter group quantum numbers.

The $AdS/CFT$ correspondence provides a fixed relation
between the anomalous dimension of the various fields at the
boundary and the masses of the bulk states.
A result of our computations is that the requirement for
the anomalous dimensions to be rational implies that one must look for
dual KK states having  also rational masses .

The virtue of KK harmonic analysis on a coset space \cite{libro} hinges on the
possibility of reducing the computation of the mass eigenvalues of
the various kinetic differential operators to a completely algebraic
problem.
Harmonics are identified by $G$ quantum numbers, and they are acted
upon by derivatives that are reduced to algebraic operators.
Such elegant technique can be quite cumbersome for complicated cosets,
but it is quite straightforward for the simple $T^{11}$ manifold.
Indeed, it allows to go beyond the computation of the scalar laplacian eigenvalues \cite{G},
or of specific sectors of the mass spectrum \cite{JRD}.

After diagonalising different operators for
fields of various spin, we have found that
all the masses have a fixed dependence on the scalar laplacian eigenvalue
\beq
H_0(j,l,r)=6[j(j+1)+l(l+1)-1/8 r^2]
\label{accazero}
\eeq
where $(j,l,r)$ refer to the $SU(2)^2$ and $R$--symmetry quantum numbers.
This is due to the fact that on a rank one coset we have only one
functionally independent Laplace--Beltrami operator.

The full analysis \cite{CDD} reveals that the supergravity theory has one graviton
multiplet with conformal dimensions
\beq
\Delta = 1+\sqrt{H_0(j,l,r)+4},
\eeq
 four gravitino multiplets with
\beq
\Delta = -1/2 + \sqrt{H_0(j,l,r\pm 1) + 4}, \quad \Delta = 5/2 + \sqrt{H_0(j,l,r\pm 1) + 4},
\eeq
and four vector multiplets, with
\bea
\Delta = -2 + \sqrt{H_0(j,l,r)+4}, &&\nonumber\\
\Delta = 4 + \sqrt{H_0(j,l,r)+4}, &&\\
 \Delta = 1 + \sqrt{H_0(j,l,r\pm 2)+4}. && \nonumber
\eea

The above formulae clearly show that rational values of the
conformal dimensions occur when the square roots
assume rational values
\beq
H_0+4 \in Q^2.
\eeq

This equation  is found to admit some special solutions for
\bea
j&=l=|r/2|,\\
j&=l-1=|r/2| \qquad {\rm or} \qquad l=j-1=|r/2| .
\eea
At this point we have some strong constraints on the possible
$SU(2,2|1)$ quantum numbers as well as on the $SU(2) \times SU(2)$
ones.
It is therefore an easy task to build the conformal operators
satisfying such constraints and find the corresponding bulk
supermultiplets.

While referring to \cite{CDDF} for all details,
we now show some interesting examples.

The chiral operators of the conformal field theory are given by
\bea
\label{prim}
S^k &=& Tr (AB)^k \\
\label{sec}
\Phi^k &=& Tr \left[ W^2 (AB)^k \right]\\
\label{ter}
T^k &=& Tr \left[ W_\alpha (AB)^k \right]
\eea
and are shown to correspond to hyper--multiplets containing massive recursions
of the dilaton or the internal metric (\ref{prim} and \ref{sec})
or to tensor multiplets (\ref{ter}).

More interesting are the towers of operators associated to the
semi--conserved currents.
Some of them are given by the following operators
\bea
{J}_{\a\da}^k &=& Tr(W_\alpha e^V \bar{W}_{\dot{\alpha}}e^{-V}(AB)^k), \\
{J}^k &=& Tr(Ae^V \bar{A} e^{-V} (AB)^k),
\eea
which lead  to short multiplets whose highest state is
a spin 2 and spin 1 field respectively, with masses given by
\beq
M_{J_{\alpha\dot{\alpha}}^k} = \sqrt{\frac{3}{2} k \left( \frac{3}{2} k +
4\right)}, \quad \hbox{ and } \qquad
\displaystyle
M_{J^k} = \sqrt{\frac{3}{2} k \left( \frac{3}{2} k +
2\right)}.
\eeq
These bulk states correspond to massive recursion of the graviton
and of the gauge bosons of the matter groups.

It has been explained that under certain conditions the
semi--conserved superfields can become conserved, and this is indeed the
case.
If we set $k=0$ we retrieve the conserved currents related
to the stress--energy tensor and the matter isometries .
In fact $M_{J_{\alpha\dot{\alpha}}^0} = M_{J^0} = 0$ are the
massless graviton and gauge bosons of the supergravity
theory.

We have now checked the correspondence as far as what the conformal
field theory predicts on the bulk states, but what can we learn {\sl
on the CFT}
from the analysis of the supergravity states?

\subsection{AdS $\to$ CFT}

There are essentially two aspects of the supergravity theory which
can give us new insight in the dual CFT.
The first is the existence of the so--called Betti multiplets \cite{Betti},
which give rise to additional symmetries of the boundary theory,
and the other is the presence of long multiplets with rational
scaling dimensions, which could provide us with new
non--renormalization theorems at least in the large $N$, $g_s N$
limit.
Let us now turn to the first aspect.
The non--trivial $b_2$ and $b_3$ numbers of the $T^{11}$ manifold
imply the existence of closed non--exact 2--form $Y_{ab}$ and 3-form $Y_{abc}$.
These forms must be singlets under the full isometry group, and thus
they signal the presence of new additional massless states in the
theory than those connected to the $SU(2) \times SU(2) \times
U_R(1)$ isometry.

From the KK expansion of the complex rank 2 $A_{MN}$ and real rank
4 $A_{MNPQ}$ tensors of type IIB supergravity we learn that we
should find  in the spectrum a massless vector (from $A_{\mu
abc}$), a massless tensor (from $A_{\mu\nu ab}$) and two massless
scalars (from $A_{ab}$).
This implies the existence of the so called Betti vector, tensor
and hyper--multiplets, the last two being a peculiar feature of
the $AdS_5$ compactification.
The additional massless vector can be seen to be the massless
gauge boson of an additional $U_B(1)$ symmetry of the theory.

From the boundary point of view we need now to find an
operator counterpart for such a vector multiplet and seek an
interpretation of the additional symmetry.
The task of finding the conformal operator is very easy, once we
take into account that it must be a singlet of the full isometry
group and must have $\Delta = 3$.
The only operator we can write is \cite{CDDF,KW2}
\beq
{\cal U}=Tr\ A e^V \bar A e^{-V} - Tr\ B e^V \bar B e^{-V} \qquad
(D^2 {\cal U}=\bar D^2 {\cal U}=0),
\eeq
which represents the conserved current of a baryon symmetry of
the boundary theory under which the $A$ and $B$ field transform
with opposite phase.
We have shown that the occurrence of such Betti multiplets is
indeed due to the existence of non--trivial two and three--cycles
on the $T^{11}$ manifold.
This implies that, from the stringy point of view, we can wrap the
$D3$--branes of type IIB superstring theory around such 3--cycles
and the wrapping number coincides with  the baryon number of
the low--energy CFT \cite{KW2}.

We finish by commenting on the second AdS prediction on the CFT.
We have shown that the conformal operators with protected
dimension are given by chiral ones or by their products with the conserved
currents.
The surprising output of the supergravity analysis
is that there exist some long operators (not protected by supersymmetry)
which have rational conformal dimension.

If we take for example the chiral operator $Tr( W^2 (AB)^k)$, we
can make it non--chiral by simply inserting into the trace an
antichiral combintation of the gauge field--strength $Tr( W^2 e^V \bar{W}^2 e^{-V}
(AB)^k)$.
This operator then corresponds to a long multiplet in the bulk
theory and one should expect its scaling dimension to be
renormalized to an irrational number.
If we search for the corresponding vector multiplet in the
supergravity theory, we see that its anomalous dimension is instead
rational and matches exactly the naive sum of the dimensions of
the operators inside the trace.
From our analysis this appears to be the case for all the lowest
non--chiral operators of general towers with irrational scaling
dimension.
For instance, the  towers of operators
\bea
Tr\left[ W_\alpha (A e^V \bar A e^{-V})^n (AB)^k\right] \\
Tr\left[ e^V \bar W_{\dot{\alpha}} e^{-V} (A e^V \bar A e^{-V})^n (AB)^k\right]
\eea
have an irrational value of $\Delta$ for generic $n$, but
when $n = 1$ we have found that they do have rational anomalous
dimension $\Delta = 5/2 + 3/2 k$.
When $n=0$ we retrieve the chiral, or semi--conserved operators
with protected $\Delta$.

\ack{
S. F. is supported in part by
the DOE under grant DE-FG03-91ER40662, Task C, the NSF grant
PHY94-07194. This work is also supported by the ECC TMR
project ERBFMRXCT96-0045 (Politecnico of Torino and Frascati)
and by INFN, Sezione di Torino and Frascati.}

\end{document}